\documentclass[%
 reprint,
 amsmath,amssymb,
 aps,
]{revtex4-1}

\usepackage{graphicx}
\usepackage{dcolumn}
\usepackage{bm}
\usepackage{physics}

\graphicspath{{Figures/}}

\begin{document}

\preprint{APS/123-QED}

\title{Exploring itinerant states in divalent hexaborides using rare-earth $L$ edge resonant inelastic X-ray scattering}

\author{Donal Sheets$^{1,2}$}
\author{Vincent Flynn$^{1,3}$}
\author{Jungho Kim$^{4}$}
\author{Mary Upton$^{4}$}
\author{Diego Casa$^{4}$}
\author{Thomas Gog$^{4}$}
\author{Zachary Fisk$^{5}$}
\author{Maxim Dzero$^{6}$}
\author{Priscilla F. S. Rosa$^{7}$}
\author{Daniel G. Mazzone$^{8}$}
\author{Ignace Jarrige$^{8}$}
\author{Jian-Xin Zhu$^{7}$}
\author{Jason Hancock$^{1,2}$}

\affiliation{$^{1}$Department of Physics, University of Connecticut, Storrs, CT 06269, USA}
\affiliation{$^{2}$Institute of Material Science, University of Connecticut, Storrs, CT 06269, USA}
\affiliation{$^{3}$Department of Physics and Astronomy, Dartmouth College, Hanover, NH 03755, USA}
\affiliation{$^{4}$Advanced Photon Source, Argonne National Laboratory, Argonne, Illinois 60439, USA}
\affiliation{$^{5}$Department of Physics, University of California, Irvine, CA 92697, USA}
\affiliation{$^{6}$Department of Physics, Kent State University, Kent, OH 44240, USA}
\affiliation{$^{7}$Los Alamos National Laboratory, New Mexico 87545,USA}
\affiliation{$^{8}$National Synchrotron Light Source II, Brookhaven National Laboratory, Upton, NY 11973, USA}

\date{\today}

\begin{abstract}
We present a study of resonant inelastic X-ray scattering (RIXS) spectra collected at the rare-earth $L$ edges of divalent hexaborides YbB$_6$ and EuB$_6$. In both systems, RIXS-active features are observed at two distinct resonances separated by $\sim10$ eV in incident energy, with angle-dependence suggestive of distinct photon scattering processes. RIXS spectra collected at the divalent absorption peak strongly resemble the \textit{unoccupied} 5$d$ density of states calculated using density functional theory, an occurrence we ascribe to transitions between weakly-dispersing 4$f$ and strongly dispersing 5$d$ states. In addition, anomalous resonant scattering is observed at higher incident energy, where no corresponding absorption feature is present. Our results suggest the far-reaching utility of $L$-edge RIXS in determining the itinerant-state properties of $f$-electron materials.


\end{abstract}

\pacs{Valid PACS appear here}
\maketitle


\section{\label{sec:level1}Introduction}

Resonant inelastic x-ray scattering (RIXS) is an emerging technique capable of probing electronic structure and collective modes of crystals and films of correlated electron systems. Using resonance enhancement of electronic transitions near X-ray edges, this photon-in/photon-out scattering technique permits exploration of collective modes where energy, momentum and polarization changes of the scattered photon deliver vital information on the energetics and quantum numbers of complex materials \cite{Ament2010}. Recent prominent applications of RIXS mainly focus on transition metal oxides and include the direct observation of charge density waves in high-$T_c$ superconductors \cite{Ghiringhelli2012}, Mott gap dispersion in their parent compounds \cite{Abbamonte1999, Hasan2000, Chabot-Couture2010}, residual paramagnon fluctuations in doped cuprates \cite{LeTacon2011}, and detailed measurement of magnetic and electronic structure in thin films \cite{Dean2012} and interfaces \cite{Zhou2011}. Sustained synchrotron endstation developments have permitted an expanded inquiry into the spin and charge spectra of a broader class of materials which harbor interesting physical phenomena.

Rare earth hexaborides form in the cubic CsCl type lattice structure and exhibit diverse, interesting, and well-characterized electronic and magnetic ground states such as putative topological Kondo insulating state in SmB$_6$ \cite{Dzero2016}, superconductivity in LaB$_6$ \cite{Schell1982},  hidden quadropolar order in CeB$_6$ \cite{Koitzsch2016}, and colossal magnetoresistance in ferromagnetic EuB$_6$ \cite{Manna2014}. In order to address the systematic variations among this class of materials which underpin their ground states, we focus our attention on two representative divalent systems: nonmagnetic YbB$_6$ and ferromagnetic EuB$_6$. YbB$_6$ is a non-magnetic small-gap semiconductor \cite{Tarascon1980}, whereas EuB$_6$ is a magnetic semiconductor with a ferromagnetic ordering transition at 15.5 K \cite{Blomberg1995,Wigger,Kim2013,Sullow1998,Kang2016a,Blomberg1995}. In each case, the covalent bonds among B atoms forms an octahedral cage with a complex valence structure and the $R$ ions sit at sites of cubic symmetry \cite{Sullow1998, Kimura1992}. Rare earth $L$-edge RIXS investigations are relatively uncommon but recent success in detecting a quasi-gap and dramatic band reconstruction at the valence transition in YbInCu$_4$ \cite{Jarrige2015} as well as recent activity assessing the topological status of YbB$_6$ \cite{Kimura1992,Xia2014a,Xu2014a,Zhou2015,Kang2016a,Neupane2015,Ramankutty2016} has motivated our systematic investigation presented here.

 \begin{figure}[t]
	\includegraphics[width=.46\textwidth, clip =true, trim = 5 30 520 25]{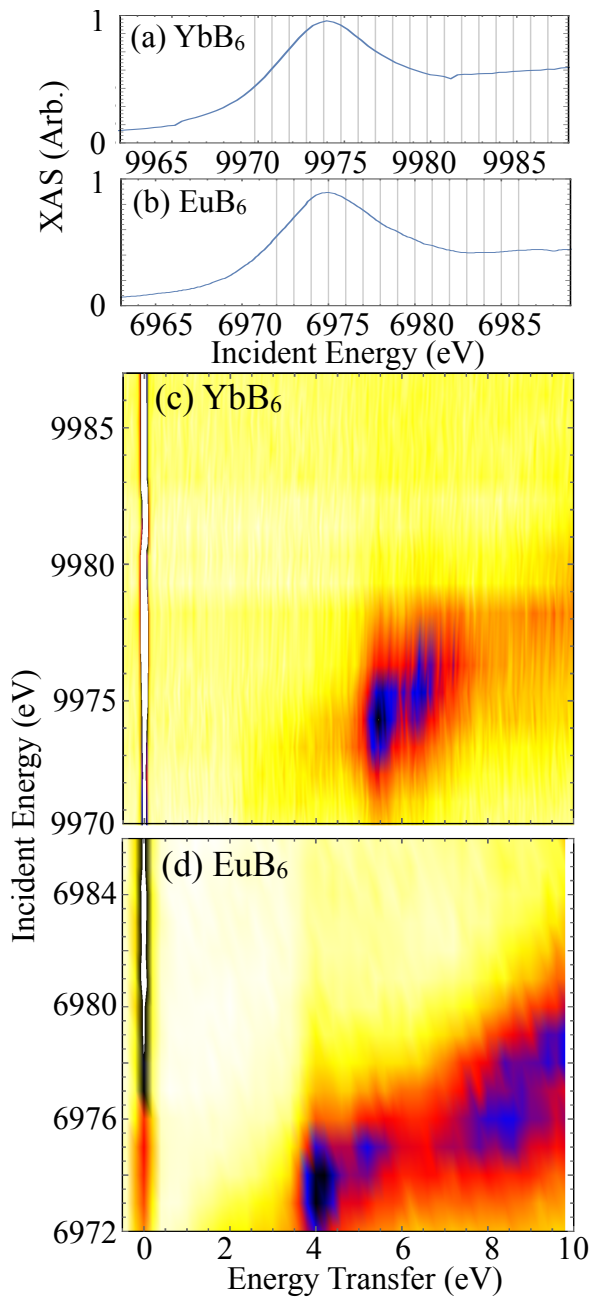}	
	\caption{X-ray absorption spectrum for (a) the $L_2$ edge of YbB$_6$ and (b) the $L_3$ edge of EuB$_6$. The vertical lines indicate the incident energies selected for the RIXS spectra below. (c) shows the RIXS spectrum for YbB$_6$ as a function of incident photon energy and energy transfer in the depolarized scattering geometry. (d) RIXS spectrum for EuB$_6$ in the depolarized scattering geometry.}
	\label{fig:contour&XAS}
\end{figure}

\section{\label{sec:level1}Experimental details}

RIXS measurements were performed at the MERIX spectrometer at the Advanced Photon Source at Argonne National Lab \cite{Shvydko2013}.  Here we investigate the $L_2$ edge of YbB$_6$ and $L_3$ edge of EuB$_6$ while varying incident energy and polarization conditions, culminating in a rather exhaustive report of the RIXS scattering opportunities in divalent rare earth materials. Undulator magnets produce horizontally polarized photons whose bandwidth are then narrowed using two pairs of monochromator crystals. Two Kirkpatrick-Baez mirrors deliver a focused 40 $\times$ 10 $\mu$m  spot on the sample. Inelastically scattered photons emerging from the scattering volume are collected by a spherical diced crystal analyzer cut from a silicon substrate on a 2m arm and backscattered onto a position sensitive micro-strip detector. Overall, energy resolutions of 75 meV ($L_2$ YbB$_6$) or 140meV ($L_3$ EuB$_6$) were achieved. The YbB$_6$ single crystals where grown by the Al-Flux method as described previously \cite{Fisk1989}.

The MERIX spectrometer \cite{Shvydko2013} has three independent axes of rotation on the sample position as well as two axes for the analyzer/detector arm. This flexible geometry allows both horizontal and vertical scattering planes, permitting both angle and energy resolution of the RIXS spectrum and the transverse polarization of the incident photons permits polarization analysis in the scattering process \cite{Shvydko2013}. The momentum transfer is given by $\vec{Q}$ = $\vec{k}_i$-$\vec{k}_f$ where $\vec{k}_i$ is the incident photons and $\vec{k}_f$ is the scattered photon, indexed by the Miller indices $(H K L)$ of the simple cubic lattice with a seven-atom unit cell. 

X-ray absorption spectroscopy (XAS) was collected in total fluorescence yield mode with an energy-integrating detector at 90 degrees from the sample to reduce the elastically scattered background while the incident energy from an attenuated beam directly from the monochromator is scanned.

\section{\label{sec:level1}Experimental Results at the Divalent Resonance}

Figure \ref{fig:contour&XAS} compares the X-ray absorption spectra (XAS) at the $L_2$ (2$p_{1/2}\rightarrow 5d$) edge of YbB$_6$ and the $L_3$ (2$p_{3/2}\rightarrow 5d$) edge of EuB$_6$. In each case we observe a similar prominent divalent peak and a complete lack of trivalent peak, expected about 7 eV higher in incident energy as observed in mixed rare earth systems \cite{Zhou2015,Jarrige2015, Jarrige2013}, confirming the divalency of each compound.  Consistent with the magnetic response of each material, the XAS data support the divalency of YbB$_6$ ($f^{14}$) and EuB$_6$ ($f^7$) and imply that each have zero orbital angular momentum associated with the $f$-electron degree of freedom.

\begin{figure}
	\includegraphics[width=.48\textwidth, clip =true,trim = 5 25 515 40]{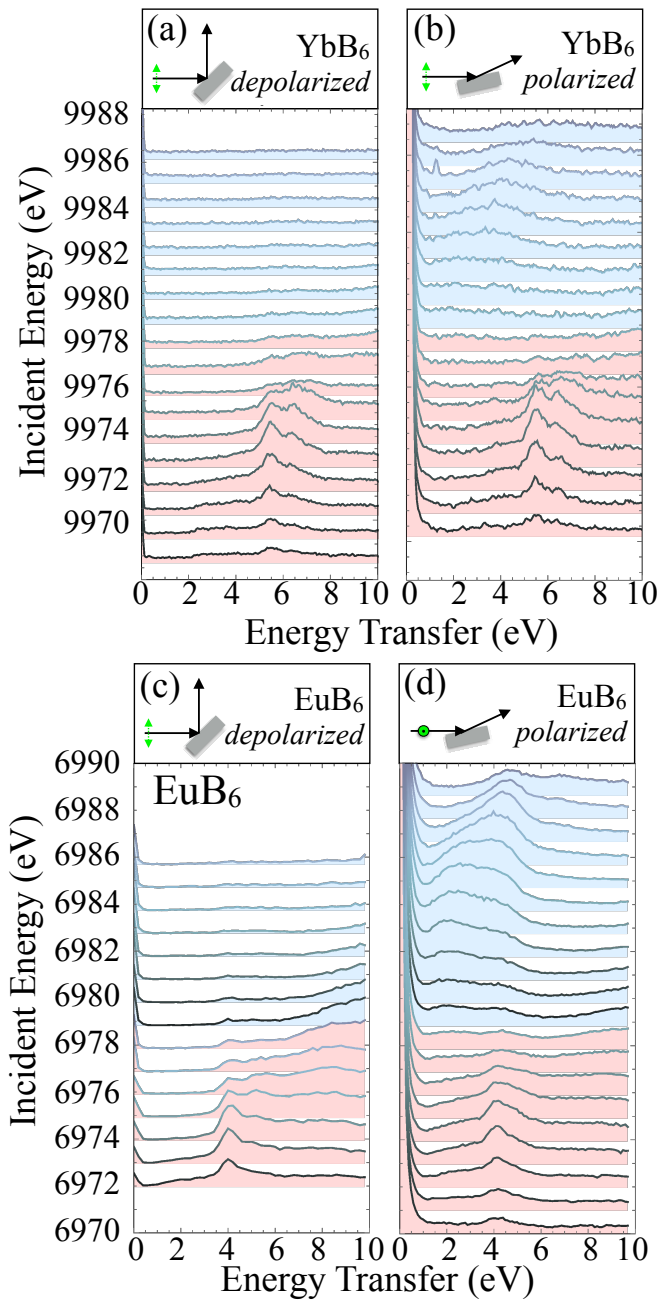}
	
	\caption[]{ RIXS data for YbB$_6$ and EuB$_6$ in the depolarized (a, c) and polarized (b, d) scattering geometries at $T = 300$ K. The insets show the incoming photon polarization as a green arrow while the incident and scattered photon momenta are indicated by black arrows. Pink and blue shading indicate respectively the low and high incident photon energy.}
	\label{fig:RIXScomparison}
\end{figure}

The RIXS spectra for the $L_2$ edge of YbB$_6$ and $L_3$ edge of EuB$_6$ at incident energies represented in Fig. \ref{fig:contour&XAS}a,b is shown in Fig. \ref{fig:contour&XAS}c,d as a color plot and Fig. \ref{fig:RIXScomparison}a,c as line scans. These spectra were collected in the $depolarized$ scattering geometry with momentum transfer $\vec{Q}$ = (4.75, 0, 0) at $T = 300$ K. In this geometry, the polarization of the scattered photon on the incident one is necessarily zero, a point we return to later. Further, this geometry minimizes the nonresonant elastic (Thomson) scattering, permitting resolution of low-energy inelastic features.

\begin{figure}
	\includegraphics[width=.5\textwidth, clip =true, trim = 5 10 285 25]{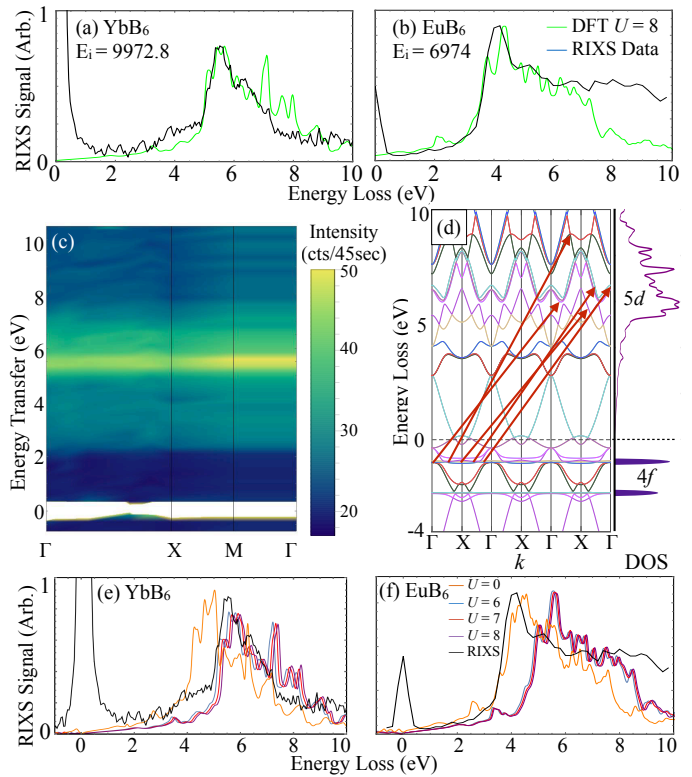}
	
	\caption{Comparison of RIXS spectrum with $5d$ density of states calculations with $U$ = 8 for (a) YbB$_6$ and (b) EuB$_6$. The RIXS spectra are shown in black with the DFT-calculated $5d$ density of states in green. In (a) the YbB$_6$ DOS was downshifted by .3 eV, in (b) the EuB$_6$ DOS was downshifted by 1.2 eV. (c) the momentum dependence of the low energy feature in YbB$_6$ along the $\Gamma$-X-M-$\Gamma$ cut, showing weak dispersion. (d) schematic of a set of net $4f$ to $5d$ transitions at constant momentum transfer from flat $4f$ bands to unoccupied $5d$ states. (e) and (f) comparison of unshifted DFT calculation for $U$ = 0,6,7,8 eV and RIXS spectra. In (e, f) no shift was applied to the DOS.
	}
	\label{fig:theorycomparison}
\end{figure}

Both materials show broad features which resonate at the divalent XAS peak and in both cases we identify a fluorescence tail above 8eV energy transfer. The rare-earth $L$ RIXS process can be considered as resonant Raman scattering dominated by dipole-allowed transitions ($\Delta l \pm 1)$ in a $2p\rightarrow 5d\rightarrow 2p$ scattering event with a strong 2$p$ core hole potential in the intermediate state. The final state, however, does not contain a core hole and the accessible range of net excitation energies persists down to the instrumental resolution $<$ 100meV. Such low energy scales can be thermally populated at ambient conditions and are therefore relevant for material behavior. For these reasons, rare earth $L$ edge RIXS holds great promise to exploring collective modes in correlated electron systems.

In order to establish detailed connections between the RIXS spectrum and underlying electronic structure, we have performed electronic structure calculations using the full-potential linearized augmented plane wave (FP-LAPW) method implemented in the WIEN2k package \cite{Blaha2001}. The generalized gradient approximation (GGA) \cite{Perdew1996} was used for the exchange-correlation functional.  The spin-orbit coupling was included in a second variational way. The muffin-tin radius $2.5a_0$ ($a_0$ being the Bohr radius), $2.5a_0$, and $1.57a_0$ and $1.60a_0$ for Yb, Eu, and B in YbB$_6$ and EuB$_6$, respectively, and a plane-wave cutoff RKmax = 8 was taken. Hereafter all theoretical calculations were performed at the experimentally determined lattice constants \cite{Blomberg1995,Bolgar1993}. To address the electronic correlation effects on Yb- and Eu-4$f$ electrons, we have also performed the GGA+$U$ calculations by using the self-interaction correction double-counting scheme \cite{Anisimov1993}. We varied the value of the Hubbard repulsion  as $U_{\text eff}$ = 6, 7, 8 eV.  The above choice of Hubbard $U$ values are on par with those reported in earlier work \cite{Kunes,Ghosh2004}. Spin-polarized GGA calculations gives vanishing magnetic moment for YbB$_6$ and $\sim 7\; \mu_{B}$ for EuB$_6$, in reasonably good agreement with the experimental measurements \cite{Tarascon1981,Fisk1979,Sullow1998} and earlier calculations \cite{Ghosh2004}.

Although the nonmagnetic state of YbB$_6$ or the moment size of EuB$_6$ are robust against the value of the Hubbard repulsion, the position of 4$f$-orbital energy levels are pushed away below the Fermi energy by this repulsion. In particular, the obtained Yb-4$f$ level positions for $U_{\text eff}=8$ eV are in good agreement with the ARPES measurements on YbB$_6$ \cite{Neupane}. Therefore, we present the electronic structure results for this specific value.

Figures \ref{fig:theorycomparison}a,b compares the RIXS spectra with the corresponding calculation of the 5$d$ density of states (DOS), sampled on a mesh of $15\times15\times15$ k points. The DOS have been shifted in energy to enable detailed comparison of the RIXS line shape and the calculated density of states, including details like the gradual rise below the strong cusp due to a prominent van Hove singularity. The calculation appears to produce sharper features at higher energies, which are likely blurred in the data by a combination of instrumental resolution and coupling to lattice degrees of freedom, as is well-known in optical and RIXS spectra \cite{Hancock2009,Hancock2010}. The observed agreement strongly supports an interpretation where electrons are excited into 5$d$ final states from a band of weakly dispersing initial states. We propose here a scattering process wherein electrons originating in the weakly dispersing 4$f$ states are scattering by the core hole potential to 5$d$ final states, illustrated schematically in Fig.\ \ref{fig:theorycomparison}d. Because the $f$ band is both occupied and weakly dispersive, the manifold of all such transitions at the momentum set by the scattering geometry provides an image of the unoccupied 5$d$ density of states. 

Figure \ \ref{fig:theorycomparison}c presents RIXS spectra collected along high-symmetry cuts in the simple cubic Brillouin zone of YbB$_6$ at the divalent resonance. Weak dispersion is apparent and supports our interpretation of the scattering process considering that the 4$f$ manifold is completely full and the 5$d$-derived states are nearly empty \cite{Neupane}. A similar momentum-independence was observed in EuB$_6$, where similar conditions are present, although in this case, the 4$f$ states are only half full. This situation presents strong contrast to RIXS studies on 5$d$ elements such as Ir and Os, which often display local moment magnetism arising from correlated states within the 5$d$ manifold \cite{Calder2016}.

Given our interpretation of the rare-earth $L$ egde RIXS spectra collected at the divalent resonance, we consider here the implications for the electronic structure of each material.  The van Hove singularity in YbB$_6$ giving rise to the strong cusp is separated energetically from the 4$f$ level by approximately 1.5eV more than in EuB$_6$. We also note that the bandwidths obtained here from the RIXS data and from the DFT calculation are in good agreement without the need to scale, a step that is often invoked to account for correlation effects \cite{Liu2015,Kobayashi2015}, which appear to be absent or muted in the unoccupied 5$d$ states of the divalent hexaborides EuB$_6$ and YbB$_6$. Such detailed information on electronic structure with orbital selectivity can serve as a benchmark in the development of new computational schemes in correlated electron systems.

\section{\label{sec:level1}Geometrical dependence of RIXS processes}

\begin{figure}
	\includegraphics[width=.5\textwidth, clip =true, trim = 40 50 350 50]{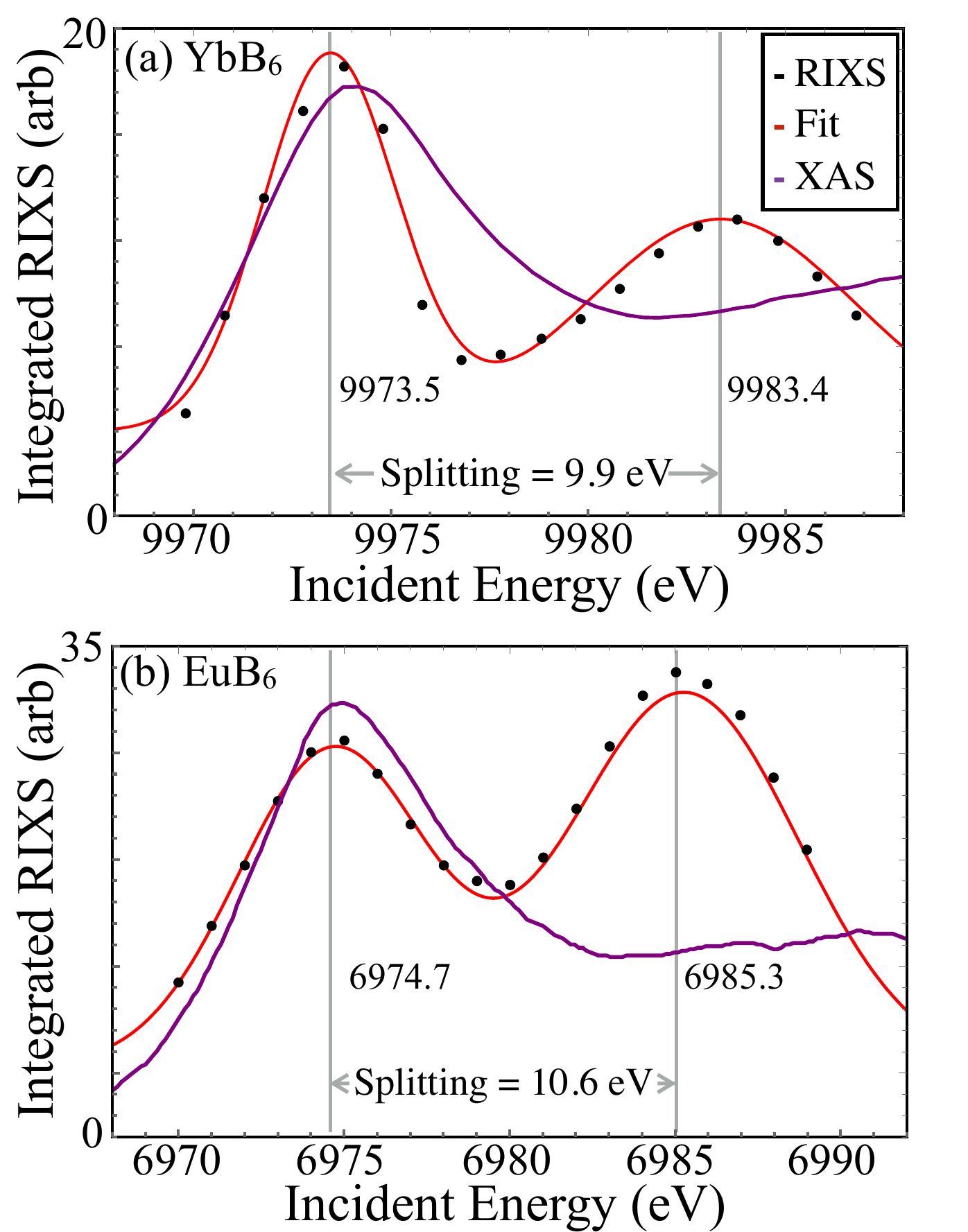}
	
	\caption{Total integrated RIXS as a function of incident energy for (a) YbB$_6$ and (b) EuB$_6$. Two peaks are apparent and split by 9.9 eV for YbB$_6$ and 10.6 eV for EuB$_6$. The lower energy peak resonates near the peak in XAS, while the higher energy feature does not have a corresponding peak in the XAS.   }
	\label{fig:IntegratedRIXS}
\end{figure}

So far we have considered the RIXS spectra of YbB$_6$ and EuB$_6$ near the divalent resonance taken in the depolarized scattering geometry with scattering angle 2$\theta$$\simeq 90^\circ$. In this configuration, the projection of the scattered photon polarization on the incident one is necessarily zero, constraining the allowed matrix elements and therefore types of excitations observed. We now relax this stringent condition by surveying a polarized scattering condition, illustrated in Fig. \ref{fig:RIXScomparison}b,d, collected at $\vec{Q}$ = (1.5,0,0) and 2$\theta$ near $25^\circ$ at  $T = 300$ K. 

As with the depolarized case, we observe a feature resembling the 5$d$ density of states which is strongest at the divalent resonance. In contrast, we also observe a strong set of RIXS transitions at a somewhat higher energy above the divalent peak. For each compound, the high energy feature peaks at a low energy transfer of $\sim$ 4 eV (Fig. \ref{fig:IntegratedRIXS}c,d), with a continuum of intensity persisting down to the lowest energies measured. For the case of EuB$_6$, this scattering has been shown to be sensitive to the bulk magnetization of the system. Specifically, below the Curie temperature $T_c$=15.5 K \cite{Sullow1998} a pronounced spectral weight redistribution toward low energy is observed in the RIXS spectra collected at high incident energy. This observation demonstrates the importance of the high-energy resonance because the final states of this scattering process result in excited states which can also be thermally excited and therefore directly influence material behavior. 

Figures \ref{fig:IntegratedRIXS}a,b show the total integrated RIXS intensity as a function of incident energy. The lower peak coincides well with the divalent peak in the XAS, while the higher energy peak occurs in a region where the XAS has no pronounced feature. The two-peak structure was fit to two Gaussian profiles plus a constant background to determine the splitting is $\simeq$10eV, which is significantly greater than the documented divalent-trivalent splitting observed to be around 7eV \cite{Jarrige2015}. This hidden resonance therefore appears not to coincide with any observable peak in X-ray absorption and furthermore cannot be attributed to a remanent trivalent state.

\begin{figure}[t]
	\includegraphics[width=.48\textwidth, clip =true]{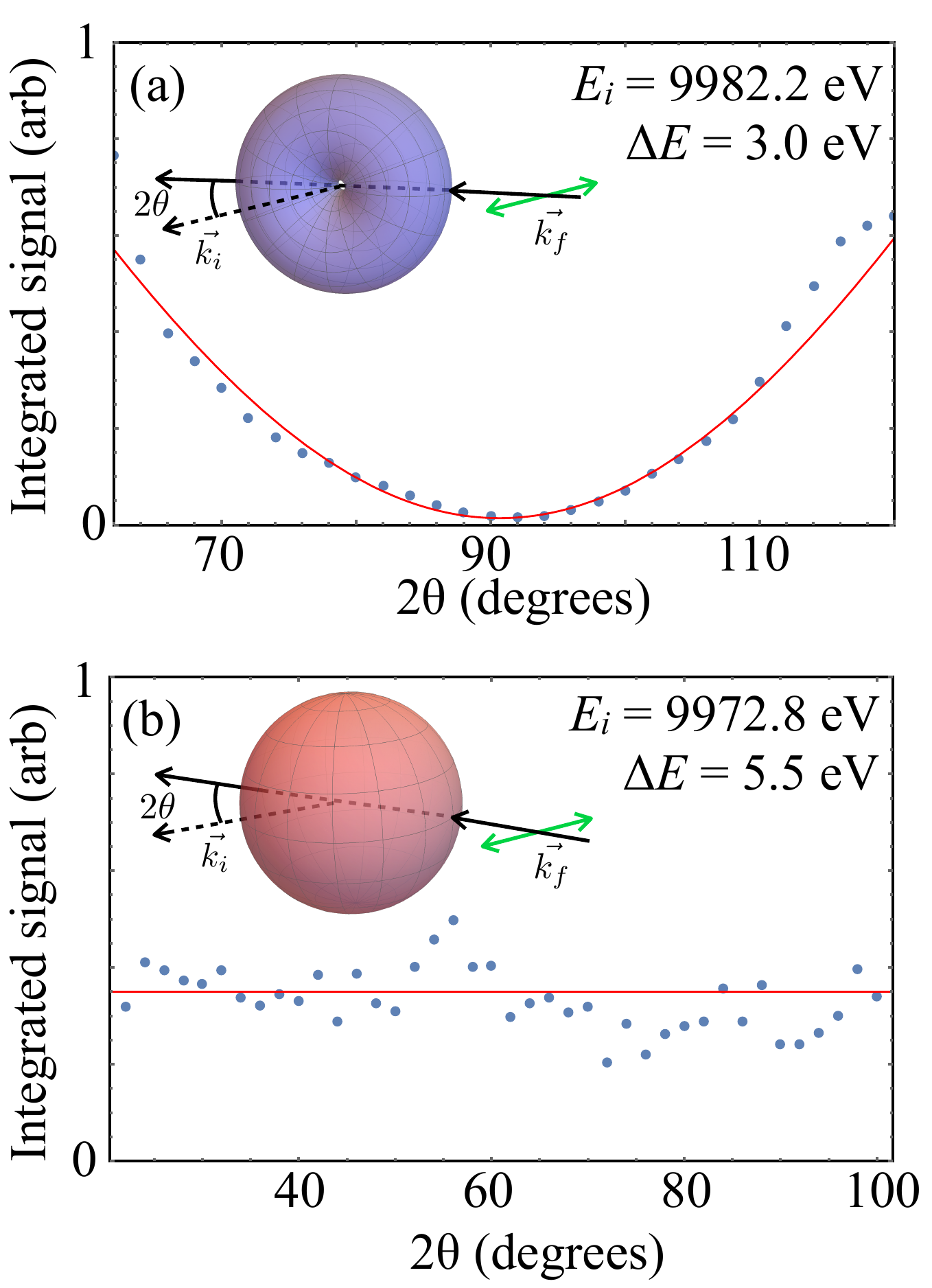}
	
	\caption{Theta-two theta scan showing the angle dependence of the two RIXS features in the spectrum of YbB$_6$. (a) shows the angular dependence of the high energy feature at $E_i = 9982.2$ eV with 1.5$<$$\Delta E$$<$4.5 eV, showing a clear node at $90^\circ$. Red solid line shows a fit to cos$^2(2\theta)$. Inset summarizes the observed polarization dependence, scattering geometry and incident polarization vector. (b) shows that the low energy feature at $E_i = 9972.8$ eV with 4 eV $<$$\Delta E $$<$7 eV has no significant angular dependence, consistent with the horizontal line shown in solid red. The isotropic scattering in this case is represented in the inset as a sphere.  }
	\label{fig:angledependence}
\end{figure}

The scattering intensity produced at high incident energies is absent in the depolarized geometry, an observation we explore and quantify here for YbB$_6$. Figure \ref{fig:angledependence}a displays a theta-two theta angular scan with $E_i$=9982.2 eV and the energy transfer integrated over a region between 1.5 and 4.5 eV to represent the RIXS spectral weight of the high energy resonance. Such a scan corresponds to a linear pathway in reciprocal space which traverses many Brillouin zones while the polarization undergoes smooth variations throughout the scan, showing clearly the behavior arises from polarization effects. Also shown is a fit to a simple angular-dependent form $\cos^22\theta$, reminiscent of Thomson scattering. Such an angular dependence has been observed before at the $K$ edge of Cu in zero-dimensional cuprate CuB$_2$O$_4$ \cite{Hancock2009}, which corresponds to an indirect scattering process, discussed below. 

The angle dependence of the 4$f$$\rightarrow$5$d$ excitation produced at the divalent resonance is shown in Fig. \ref{fig:angledependence}b, and can be well described as being totally independent of scattering angle. This scattering distribution over angle can be represented as a sphere, shown schematically in the inset of Fig. \ref{fig:angledependence}b. This angle independent scattering is consistent with an attribution of a indirect $4f$ transition to a $5d$ band with small occupation.

\section{Discussion}

In an indirect RIXS scattering process, the photo-excited electron enters electronic orbitals far above the Fermi level and plays only a small part in generating the excitations, which are dominated by the strong core hole potential. For indirect Cu $K$ RIXS, theoretical models which neglect the interaction of the 4$p$ ``spectator" electron provide adequate description of RIXS spectra in cuprates \cite{Hasan2000,Chen2010,Ament2010,Hancock2009}, which are driven by the 1$s$ core hole potential. For the rare-earth $L$ edge RIXS considered here, the 5$d$ states play the role of the spectator and 2$p$ core holes generate the excitations.

Though we have made compelling progress in interpreting the nature of the excitations generated at the divalent resonance, spectra collected at higher energy  present significant challenges. We note that the numerical value of the splitting between the peaks of RIXS is rather large: in YbB$_6$, two peaks in the integrated RIXS spectrum are separated by 9.9 eV, while for EuB$_6$ there is a splitting of 10.6 eV between the peaks. Rare-earth hexaborides have long been studied and are in use as thermionic emitters due to their low work function and high melting point. For each case presented here, the resonance splitting far exceeds the work function of each material $\sim$4 eV \cite{Wang2013}, suggestive that to the extent that the fermi level is measured by the divalent peak, the intermediate states generating the high energy peak are unbound from the material. The intermediate states do not then correspond to final states of XAS, but instead correspond to the final states of photoemission. Further work is needed to fully explore the implications of the scenario.

\section{\label{sec:level1}Conclusions}

Our resonant inelastic X-ray scattering investigation of the $L$ edges of rare-earth hexaborides YbB$_6$ and EuB$_6$ shows two strong resonant features attributable to distinct RIXS scattering processes. We present compelling evidence that an indirect RIXS feature present at the divalent peak of x-ray absorption has negligible polarization and momentum dependence is due to a net $4f \rightarrow 5d$ transition. Quantitative considerations of this feature provide microscopic energy scales such as the 5$d$ bandwidth, energy of the 4$f$ electronic bands, and systematic variation across materials. In addition, we observe a second set of features resonant at a higher incident energy where no peak in absorption can be identified. This feature has strong polarization dependence resembling that of Thompson scattering. Our results suggest far-reaching utility of $L$ edge RIXS experiments in deciphering mixed valent, heavy fermion, hidden order, and other complex phenomenology of $f$-electron materials at a microscopic level. In particular, extension of our quantitative rare-earth $L$ edge RIXS work and relation to the underlying electronic structure to more complex systems such as SmB$_6$ and YbB$_{12}$ hold strong promise toward evaluation of the topological status of these and related 3D materials.

\section{\label{sec:level1}Acknowledgements}

Work at the University of Connecticut and Kent State University was supported by the U.S. Department of Energy, Office of Science, Office of Basic Energy Sciences, under Award No. DE-SC0016481. This research used resources of the Advanced Photon Source, a U.S. Department of Energy (DOE) Office of Science User Facility operated for the DOE Office of Science by Argonne National Laboratory under Contract No. DE-AC02-06CH11357. Daniel G. Mazzone acknowledges funding from the Swiss National Science Foundation, Fellowship No. P2EZP2-175092. Work at Los Alamos National Laboratory was supported by U.S. DOE BES Core Program E3B5. Sample synthesis at Los Alamos was supported by the U.S. DOE BES program "Quantum Fluctuations in Narrow Band Systems". Work at Kent State University by Maxim Dzero was financially supported in part by NSF-DMR-1506547. 

\bibliography{REHexaboride,library}
\end{document}